\documentclass[12pt]{iopart}
\usepackage{graphicx,color,bm}      
\usepackage{hyperref}
\definecolor{clr}{rgb}{0, 0.25, 0.75}
\hypersetup{
  colorlinks=true,
  linkcolor=clr,
  citecolor=clr,
  filecolor=magenta,
  urlcolor=clr,
  pdfpagemode=UseThumbs,
  pdfstartview=FitH,
}
\usepackage{cite}

\begin{document}
\topical{Rydberg atom quantum technologies}
\author{C. S. Adams$^1$, J. D. Pritchard$^2$, J. P. Shaffer$^3$}
\address{$^1$ Department of Physics, Durham University,
Rochester Building, South Road, Durham DH1 3LE, UK}
\address{$^2$ Department of Physics, University of Strathclyde,
John Anderson Building, 107 Rottenrow East, Glasgow G4 0NG, UK}
\address{$^3$ Quantum Valley Ideas Laboratories, 485 West Graham Way, Waterloo, ON N2L 0A7, Canada}
\ead{jshaffer@qvil.ca}

\begin{abstract}
This topical review addresses how Rydberg atoms can serve as building blocks for emerging quantum technologies. Whereas the fabrication of large numbers of artificial quantum systems with the uniformity required for the most attractive applications is difficult if not impossible, atoms provide stable quantum systems which, for the same species and isotope, are all identical. Whilst atomic ground-states provide scalable quantum objects, their applications are limited by the range over which their properties can be varied. In contrast, Rydberg atoms offer strong and controllable atomic interactions that can be tuned by selecting states with different principal quantum number or orbital angular momentum. In addition Rydberg atoms are comparatively long-lived, and the large number of available energy levels and their separations allow coupling to electromagnetic fields spanning over 6 orders of magnitude in frequency. These features make Rydberg atoms highly desirable for developing new quantum technologies. After giving a brief introduction to how the properties of Rydberg atoms can be tuned, we give several examples of current areas where the unique advantages of Rydberg atom systems are being exploited to enable new applications in quantum computing, electromagnetic field sensing, and quantum optics. 

\end{abstract}
\maketitle

\section{Introduction}

Recent efforts to develop quantum technology are motivated by the fact that precisely manipulating systems at a quantum level offers particular advantages over classical techniques \cite{Dowling2003}. There is widespread belief that fundamentally new types of computers, communications systems and sensors that take advantage of unique quantum properties will revolutionise society. Atoms are important building blocks for quantum technologies because, despite massive research efforts in fabrication, it is still difficult to build artificial quantum systems with the uniformity and ease required for many of the most tantalising applications. Particularly enticing for quantum technologies are Rydberg atoms---where the outer electron is placed in a highly-excited state \cite{gallagher05,sibalic18}. Rydberg atoms possess unique, exaggerated atomic properties that can be controlled by state selection and the application of external electromagnetic fields \cite{mohapatra07}, making them one of `mother nature's' most tunable quantum systems. Coupled with precision control of their properties, Rydberg atoms are close to ideal building blocks for quantum engineering.

In this review, after a brief introduction to the properties of Rydberg atoms, we discuss the current state-of-the-art in three applications areas.  First we discuss quantum simulation and computing using ultra-cold Rydberg atoms in optical tweezer arrays. Second, we discuss the sensing and imaging of microwave and terahertz fields using Rydberg atoms in room temperature vapour cells. Finally, we discuss the application of Rydberg atoms to quantum optics, including light-matter interfaces including single photon sources and photon gates. 

\section{Rydberg Atoms and Their Tunability}
\label{sec:properties}

The measurement of spectral lines by Anders Jonas
{\AA}ngstr\"om (L\"{o}gd\"o 1814 -- Uppsala 1874) revealed patterns that were later explained by Johannes (Janne) Robert Rydberg (Halmstad 1854 --
Lund 1919).  Rydberg's interval formula suggested that the energies of states of bound electrons could be written in the form
$$E_n=-\frac{R_{\rm H}}{n^2}~,$$
where $n$ is an integer known as the principal quantum number and $R_{\rm H}$ is Rydberg's constant. Subsequently, states with high $n$  became known as highly-excited Rydberg states  \cite{gallagher05,sibalic18}. These Rydberg states are interesting for a number of reasons. Firstly, because the electron is far from the nucleus and only weakly bound, it is extremely sensitive to its environment, to external electric fields or other Rydberg atoms. Also, highly-excited Rydberg states are metastable with lifetimes of order 100~$\mu$s, four orders of magnitude longer than low-lying excited states  \cite{gallagher05}. This combination means that Rydberg atoms are sensitive probes of their environment. In particular, Rydberg atoms exhibit an extreme sensitivity to either nearby Rydberg atoms or to fields in the microwave and terahertz region of the electromagnetic spectrum. These two sensitivities lead to at least two distinct classes of applications. First, the strong interactions between nearby Rydberg atoms make them attractive candidates for quantum simulators or quantum computing \cite{saffman10}, and for quantum non-linear optics \cite{firstenberg16}. Second the sensitivity of Rydberg atoms to microwave or terahertz fields make them ideal sensors of electromagnetic fields in these regions \cite{sedlacek12,fan15,wade16}. Building on developments in laser technology needed to prepare Rydberg atoms \cite{mohapatra07}, advances in laser cooling and trapping, real world applications of Rydberg quantum technologies has now become practical.

Rydberg atoms are attractive as building blocks for quantum technologies for a number of reasons. Atoms are stable quantum systems that are always the same, i.e. all atoms of the same species behave the same when placed in the same environment. Their properties can be determined by precision measurements and do not change in time. The ability to access highly-excited Rydberg states that are long-lived on useful experimental timescales is important because it allows the characteristics of the atom to be tuned. Many useful properties, such as the polarisability of the atom, scale strongly with $n$ \cite{gallagher05,bethe}. Conceptually, Rydberg states resemble the physics of hydrogen, so physicists have a straightforward model to help guide intuition, even for multi-Rydberg electron atoms like strontium. In this section, we describe how some of the most important properties of Rydberg atoms scale with $n$ and are relevant to the applications presented in the remaining sections of this tutorial.

The topics covered in this short tutorial depend on the strong interactions between Rydberg atoms and the use of the strong transitions that occur between Rydberg states. We therefore focus on Rydberg atom transition dipole moments, polarisability, and electron orbital size. These properties determine both the strength of Rydberg atom interactions and transition strengths. We also consider the lifetimes of the Rydberg states because the lifetime can determine the feasibility of some experiments, putting a limit on the coherence time of the Rydberg state. In fact, the strong scaling with $n$ occurs because as $n$ increases the binding energy of the electron becomes weaker. We will ignore magnetic effects, since the application of Rydberg atoms to quantum technologies predominantly relies on the core of the Rydberg atom being an electrostatic trap for a weakly bound electron which can be strongly perturbed by an applied electric field, as we implied in the Introduction. The Rydberg electron has spin which couples to a magnetic field, but its interaction with the atomic nucleus is weak because its orbital is large \cite{gallagher05}. For applications that concern us here, we take advantage of the fact that the Rydberg electron responds to an electromagnetic field more strongly than a valence state but more weakly than a bare electron.

To understand how a highly-excited Rydberg atom's properties scale with $n$ it is useful to use hydrogen as a model.
The moments of the electron position $r$, $\langle r^s \rangle$, where $s$ is a positive or negative integer, lead to an understanding of the scaling of the atomic properties with $n$ and the orbital angular momentum of the electron, $l$. If $s$ is positive, then $\langle r^s \rangle$ is set by the wave function at large $r$. On the other hand, if $s$ is negative, it is the behavior of the wave function near the core of the Rydberg atom that determines the moment. For $s$ positive, $\langle r^s \rangle$ will be largely independent of $l$ since it is the nodal structure at large $r$ that dominates the integral that gives the expectation value. For $s$ negative, lower $l$ states have a larger probability to be found near the nucleus because the centrifugal barrier `pushes' the wave function of higher $l$ states out to larger $r$. The high $l$ states are often referred to as circular states since the probability of finding the electron near the nucleus is practically negligible. The leading order scaling with $n$ for positive $s$ is $\langle r^s \rangle \propto n^{2s}$ while for negative $s<-1$ the scaling is $\langle r^s \rangle \propto n^{-3}$. $s = -1$ is a special case where $\langle r^{-1} \rangle \propto n^{-2}$.

For an alkali atom, the scaling of $\langle r^s \rangle$ with $n$ can be used with some modification. The inner electrons of an alkali atom form a closed shell, leaving a single electron in an unfilled outer shell, making the central field approximation highly accurate. Nevertheless, alkali atoms are multi-electron atoms. Low $l$ states penetrate into the core of the Rydberg atom and interact with the core electrons. The interaction of the Rydberg electron and the core electrons can be handled using quantum defect theory \cite{gallagher05}. As far as the scaling laws are concerned, we can replace $n$ with $n^{*} = n - \delta_l$, where $\delta_l$ is the quantum defect for an electron with orbital angular momentum $l$. $\delta_l$ decreases with increasing $l$ since the wave function is pushed to larger $r$ as $l$ grows in magnitude. Typically, if $l>4$ the Rydberg states of alkali atoms are hydrogenic, although this depends on the accuracy one needs for a particular application. As $l$ increases, $n^* \rightarrow n$.


\begin{table}
\begin{center}
\caption{\label{scaling}Alkali atom principal quantum number ($n$) scaling of the most important properties of Rydberg states. The $n$ dependence results from the characteristics of the Rydberg atom wavefunctions, as described in the text.}
\label{tab:nscaling}
\footnotesize
\begin{tabular}{@{}lll}
\br
Property &  Quantity & Scaling \\
\mr
Energy levels &  $E_n$  & $n^{-2}$\\
Level spacing  &  $\Delta E_n$  & $n^{-3}$\\
Radius &  $\langle r\rangle$  & $n^2$\\
Transition dipole moment ground to Rydberg states &  $\vert\langle n\ell\vert -e r\vert g\rangle\vert$  & $n^{-3/2}$\\
Radiative lifetime &  $\tau$  & $n^{3}$\\
Transition dipole moment for adjacent Rydberg states &  $\vert\langle n\ell\vert -e r\vert n\ell'\rangle\vert$  & $n^2$\\
Resonant dipole-dipole interaction coefficient &  $C_3$  & $n^4$\\
polarisability &  $\alpha$  & $n^7$\\
van der Waals interaction coefficient &  $C_6$  & $n^{11}$\\
\br
\end{tabular}\\
\end{center}
\end{table}
\normalsize

Using these concepts, the average radius of a Rydberg atom can be seen to scale  $\propto n^2$. The transition dipole between neighboring states $\mu =\langle e r \rangle \propto n^2$ and the polarisability $\alpha\propto n^7$. The $n$ dependence for some of the more important properties of Rydberg atoms are shown in Table~\ref{tab:nscaling}. As numerical examples, the average radius of the electron is $69\,$nm for the $40s_{1/2}$ state of cesium. The polarisability of the cesium $40s_{1/2}$ state is $10.7\,$MHz$\,$cm$^2\,$V$^{-2}$. Scaling laws can also provide insight into Rydberg atom interactions. The resonant dipole interaction that results from the resonant exchange of virtual photons between the atoms is proportional to $R^{-3}$, where $R$ is the internuclear separation. The resulting inter-atomic potential energy is $U(R) = C_3/R^3$. The $C_3$ coefficient is proportional to the transition dipole moment between the resonant levels squared. The resonant dipole interaction then scales as $n^4$. This $n$-scaling also holds for two Rydberg atoms polarised in a weak background electric field since the electric field creates a pair of static dipoles. A van der Waals potential, $U(R) = C_6/R^6$, arises from non-resonant exchange of virtual photons between the atoms in second order perturbation theory. Here the $n$-scaling is proportional to $n^{11}$ because $C_6$ is proportional to the transition dipole moments to the fourth power divided by the energy defects which scale as $\Delta\propto n^{-3}$. For $C_6$, a similar argument can be recognised by observing that $C_6$ can be calculated by considering the atomic polarisability, where the atomic polarisability, $\alpha\propto n^7$, and the energy separations between the Rydberg levels, $\propto n^{-3}$. $C_6$ can be obtained by integrating the dynamic polarisabilities over energy \cite{Craig}.

The primary decay processes for Rydberg states are radiative decay and blackbody decay. The lifetime of a Rydberg state is given by,
\begin{equation}
\tau_{\rm R} = \left(\frac{1}{\tau_\mathrm{r}} + \frac{1}{\tau_\mathrm{bb}}\right)^{-1},
\end{equation}
where $\tau_\mathrm{bb}$ is the inverse of the blackbody decay rate and $\tau_\mathrm{r}$ is the inverse of the spontaneous emission rate. The spontaneous emission rate is typically dominated by the highest frequency, allowed transition because of the cubic dependence of the decay rate on the transition frequency, $\Gamma_s = e^2 \mu_{ij} \omega_{ij}^3/3 \pi \epsilon_0 c^3$. Here $e$ is the electron charge, $\mu_{ij}$ is the dipole moment between states $i$ and $j$, and $\omega_{ij}$ is the transition frequency. The exception to this principle happens when $n$ is very large. In the case of large $n$, the transition frequencies become more and more similar and the radial matrix elements between the states determine the decay rates. For the case where the radiative decay rate is dominated by the highest frequency allowed transition, the lifetime can be conveniently parameterised by,
\begin{equation}
\tau_{\rm r} = \tau_0 n^{*\delta}.
\end{equation}
In this equation $\delta \approx 3$. Both $\tau_0$ and $\delta$ depend on species and $l$. Tables of $\tau_0$ and $\delta$ for the alkali species can be found in Ref.~\cite{gallagher05}.

Unlike valence states, transitions between Rydberg states can take place at frequencies which are thermally populated, due to thermal radiation from the environment. The thermal background radiation can couple Rydberg states and modify their lifetimes. The blackbody radiation induced decay rate can also be parameterised  so that it can be calculated in a straightforward manner,
\begin{equation}
\frac{1}{\tau_{\rm bb}} = \frac{A}{n^{*D} \left[\exp(B/(n^{*C} T)) - 1 \right]}.
\end{equation}
Here, $A$, $B$, $C$ and $D$ are parameters while $T$ is the environmental temperature. The parameters $A$, $B$, $C$ and $D$ for different Rydberg series can be found in Ref.~\cite{beterov09}. The $n$ dependence is not polynomial here.

\begin{figure}[tb]
\begin{center}
\includegraphics[width=13.5cm]{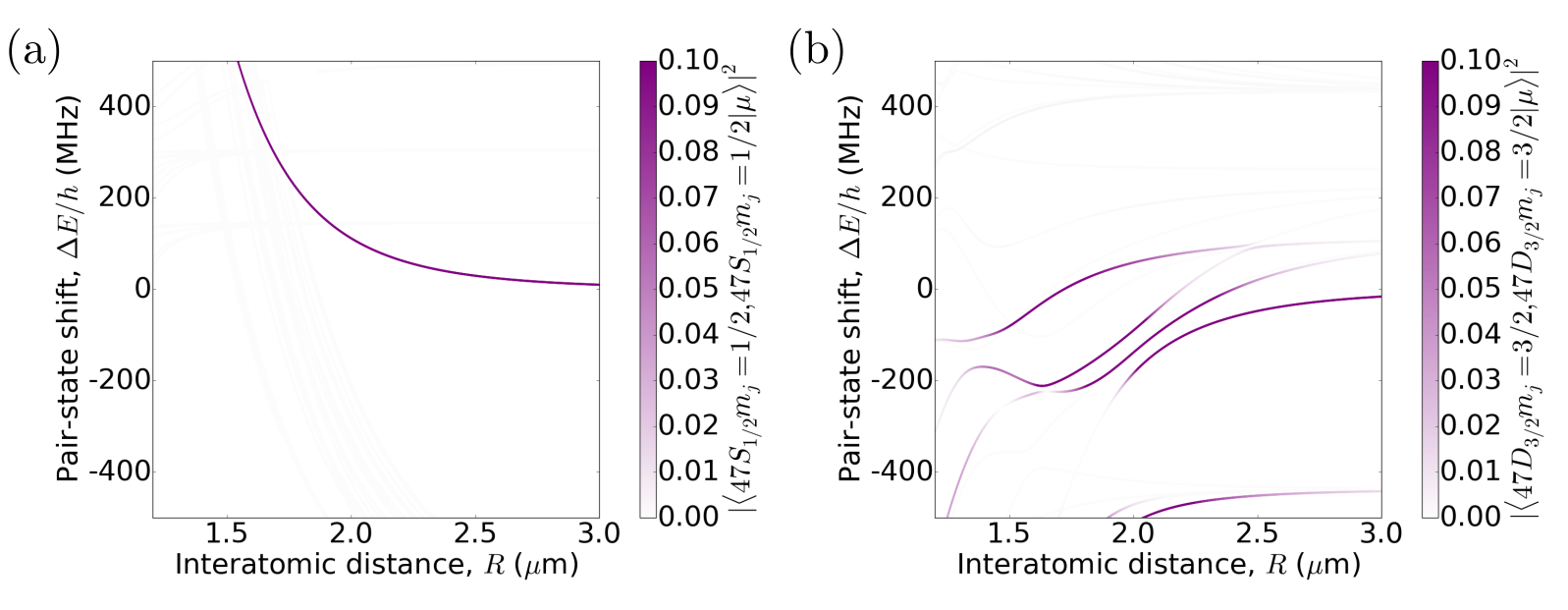}
\end{center}
\caption{Interaction potentials for pairs of Rb Rydberg atoms in the $n=47$ state with either (a) $\ell=0$ (s-states) or (b) $\ell=2$ (d-states) calculated using \texttt{ARC} \cite{sibalic17}. The frequency to excite two Rb atoms separated by 2 microns to the 47s state is shifted 100~MHz off-resonance. This interaction induced shift gives rise to Rydberg blockade. In contrast, for d-states the blockade effect is less clear-cut. }
\label{fig:pairs}
\end{figure}

As a final note, we point out that the scaling laws we have described are best used to enhance ones intuition about highly-excited Rydberg atoms and the tunability of their properties. Scaling laws are not a substitute for calculations for quantitative analysis of experiments, but they can provide guidance, particularly, since the atomic state properties are not changed independently of one another. Rydberg atom interactions \cite{marcassa2014} play a particularly important part in contemporary Rydberg atom physics and are an example of where the physics can be more complicated and deviations from simple scaling laws quite significant, particularly when electric fields are involved \cite{cabral2011}. There are publicly available calculators for Rydberg properties such as \texttt{ARC} \cite{sibalic17} or \texttt{Pairinteraction} \cite{weber17}. Many of the properties of Rydberg atoms are readily calculated if one wants to develop one's own programs. An example, for the case of Rb, is shown in Fig.~\ref{fig:pairs}. We see that states with different angular moment behave very differently, and often $s$ states with zero angular momentum are favoured due to their simple pair potentials. For example, the van der Waals interaction between n$s$ Rb atoms shifts the energy for double excitation off-resonance, such that only one excitation potential plays a role.  The distance where the excitation of further Rydberg atoms is detuned from resonance due to Rydberg atom interactions is known as the blockade radius, $R_{\rm b}=(C_6/\Delta \nu_L)^{1/6}$, where $\Delta \nu_L$ is the bandwidth of the excitation laser.

In this section, we have attempted to highlight the large range over which Rydberg atom properties can be varied in support of our assertion that these states are stable quantum systems that can be engineered for applications.

\section{Quantum Computing}

Quantum computers offer a platform to simulate quantum systems such as molecules or materials, as well as solving classically hard problems ranging from factorisation to optimisation problems such as logistics or network optimisation. Open questions surrounding the true speedup of optimisation when accounting for the resources required to encode the problem are still pending. Whilst large scale applications will require over a million qubits \cite{gorman17}, in the near term small quantum processors offer significant advantages for use in hybrid quantum-classical computation \cite{bravyi16}, also known as a quantum co-processor, with applications in quantum chemistry \cite{aspuruguzik05,lanyon10}, electronic structure calculations for complex correlated materials \cite{bauer16,barends15} as well as quantum optimal control \cite{li17} and classical optimisation problems \cite{moll18}.

Scaling of quantum systems remains a major experimental challenge, with fault-tolerant performance demonstrated for trapped ion and superconductor systems but with significant technical challenges in extending this performance to hundreds of qubits. Rydberg atoms offer a unique advantage to overcome the limitations of scalability and connectivity of competing technologies and provide a route to creating a platform of up to 1000 identical and long-coherence qubits in arrays of arbitrary geometry and dimensionality \cite{nogrette14, barredo14, barredo18}. These arrays are ideally suited for either high-fidelity digital quantum computation using the Rydberg blockade effect to engineer conditional multi-qubit gates \cite{lukin01,saffman10}, or for analog quantum computation of classical optimisation problems \cite{pichler18} and quantum simulation of complex many-body systems \cite{bernien17,lienhard18}. Here the exquisite control over the atom-atom interaction potential offered through the choice of Rydberg state and tuning using external static \cite{ravets15}, microwave \cite{maxwell13} or optical \cite{leseleuc17} electric fields allow engineering of highly anisotropic interactions with variable length scales.

\begin{figure}[tb]
\begin{center}
\includegraphics{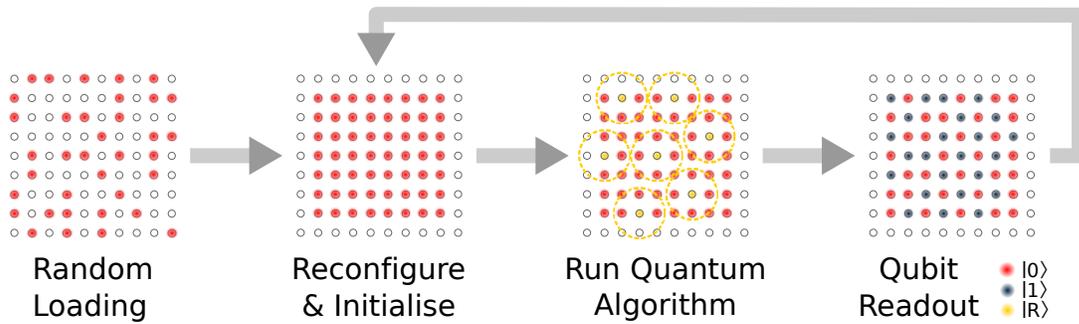}
\end{center}
\caption{Schematic of a Rydberg array quantum computer. Atoms are initially loaded stochastically, followed by rearrangement to achieve a defect free qubit register. Coherent excitation to Rydberg states allows implementation of quantum algorithms exploiting long-range interactions to couple neighbouring qubits, followed by state-selective readout which is repeated many times to tomographically reconstruct the output state. }
\label{fig:arraycomputing}
\end{figure}

Figure~\ref{fig:arraycomputing} illustrates the principal of Rydberg atom quantum computing using tweezer arrays, where atoms are loaded, and sorted before performing quantum algorithms and reading out the state of the system. Below, we consider the different regimes of digital and analog quantum computing that can be performed on this architecture.

A distinct advantage of the neutral atom platform is the ease with which a large number of identical qubits can be cooled and trapped either using optical lattices or in arrays of single-atom optical tweezer traps. Static trap configurations can be created using lenslet arrays \cite{ohl19} or fixed diffractive optical elements \cite{piotrowicz13}, whilst reconfigurable geometries are possible using either an acousto-optic deflector (AOD) \cite{lester15,endres16} or holographic projection from a spatial light modulator \cite{nogrette14} able to create complex arrays in up to 3 dimensions \cite{barredo18}. This versatility provides complete control over qubit connectivity whilst allowing dynamic programming of analog quantum simulators through choice of qubit geometry.

Single atom tweezer traps are formed using micron-scale focused beams to create a tight trapping volume that enhances atom-light interactions, leading to pairs of atoms being ejected from the trap resulting in loading of either 1 or 0 atoms in each tweezer with just over 50~\% success probability through the process of collisional blockade, as first demonstrated in 2001 by Schlosser \etal{}. \cite{schlosser01}. This probabilistic loading limits creation of large, defect-free qubit arrays through stochastic loading alone, however recently a number of techniques have been demonstrated to overcome this limitation and create deterministically assembled arrays either using dynamic trap reconfiguration using an AOD \cite{endres16} or spatial light modulator (SLM) \cite{kim16}, or via real-time sorting of atoms in 2D \cite{barredo16} and 3D \cite{barredo18} with a controllable tweezer beam to move atoms between trap sites following an initial image of a randomly loaded trap array. Alternatively, by engineering the atom-light interactions it is possible to greatly increase the probability of loading single atoms to over 90~\% \cite{carpentier13,lester15,brown19}. Combining these techniques enables reduction of trap redundancy to enable scaling to many 100s of single atom traps on timescales around two orders of magnitude faster than is possible using alternative approaches such as preparation of a Mott-insulator state within an optical lattice \cite{greiner02}. Whilst initial demonstrations of these tweezer traps have involved alkali atoms, these techniques have recently been extended to include both divalent atoms \cite{norcia18,cooper18,saskin19,jackson19} and molecules \cite{anderegg19}.

A drawback of the red-detuned tweezer traps is the strong repulsive AC Stark shift experienced by the Rydberg states, for which the polarisability is dominated by that of a free electron leading to anti-trapping. Commonly this is circumvented by turning the traps off during Rydberg excitation, leading to a finite experiment duration typically around 10~$\mu$s limited by the finite temperature of the atoms in the trap. For scalable processing longer hold times are required, which can be achieved using blue-detuned optical traps that simultaneously trap both ground and Rydberg states and can be engineered to be magic for a given transition for enhanced coherence \cite{dutta00,zhang11}. Blue-detuned trap arrays have been demonstrated using both projected traps \cite{piotrowicz13} and using a 3D optical lattice \cite{nelson07}.

\begin{figure}[tb]
\begin{center}
\includegraphics[width=15cm]{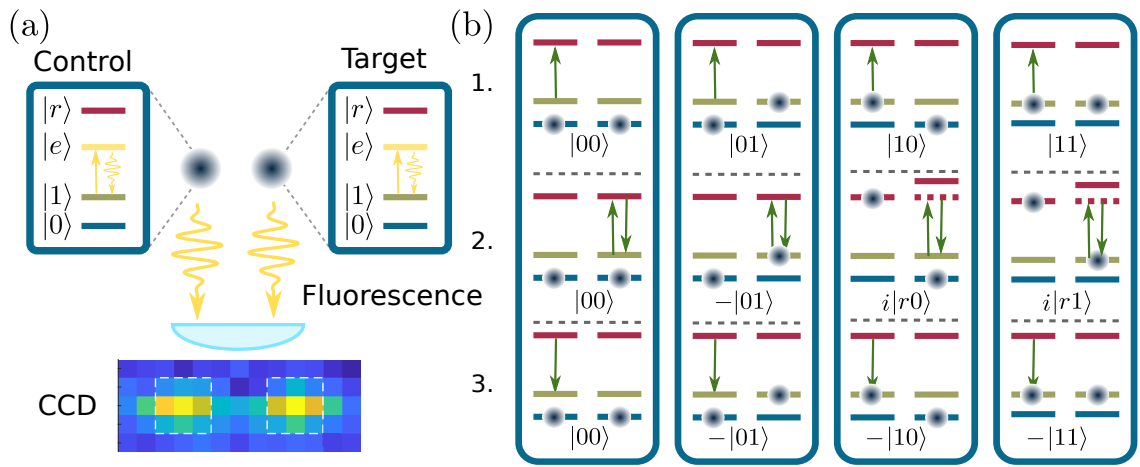}
\end{center}
\caption{(a) State-selective fluorescence for qubit-state readout using high numerical aperture lenses to collect light onto camera (b) Schematic of the controlled-phase ($C_z$) gate using Rydberg blockade to prevent both control and target atoms being in the Rydberg state simultaneously, leading to a conditional phase shift acquired on the output state.}
\label{fig:entanglement}
\end{figure}

Finally, alongside scalability important requirements for quantum information processing are the ability to perform high-fidelity state preparation and readout as well as achieving long coherence times in the computational basis. To this end initialisation of atoms into the vibrational ground state using resolved sideband cooling has been demonstrated for red-detuned tweezer traps \cite{kaufman12,thompson13,sompet17} allowing suppression of sensitivity to Doppler dephasing. Coherence can be further enhanced through magic wavelength trapping or magnetic dressing of the trap potential \cite{lundblad10,dudin13,yang16} to cancel the differential AC Stark shifts in the trap. Non-destructive state selection using an intermediate shelving state as used routinely for ions is possible for divalent atomic species \cite{norcia18,covey19}, however for the alkali's there is no easily accessible long-lived state. This has been overcome using both state-selective scattering in an optical tweezer \cite{kwon17} and spatial separation in a state-selective optical lattice \cite{wu19}, providing a route to enable rapid measurement of atomic qubits without requiring reloading of the array, as illustrated in Fig.~\ref{fig:entanglement}(a).

Digital quantum computing utilises a series of sequential gates applied onto a qubit register followed by a state-selective measurement to project atoms into the $\vert 0 \rangle$ and $\vert 1\rangle$ logical output states. For atomic qubits, information can be encoded using hyperfine ground states with a microwave-scale energy separation, or for divalent atoms using the long-lived triplet states to allow information to be encoded between ground and excited states resulting in an optical energy spacing. A complete set of single-qubit rotations can be applied by driving the two-level system using microwaves for hyperfine qubits, or optically to allow single-site resolution allowing local pulses on individual qubits, with evolution controlled by the amplitude, detuning and phase of the applied control field. To leverage the power of quantum computing to permit speed-up however, it is necessary to extend this set of single-qubit rotations to include two-qubit gates that allow entanglement generation between pairs of atoms. Using this universal gate set, all quantum algorithms can then be decomposed onto single and two-qubit gates, creating a fully programmable universal quantum computer \cite{nielsen05}.

Neutral atoms experience extremely weak interactions in the ground states, providing excellent ground state coherence times but making two-qubit gate entanglement challenging unless traps are merged to exploit collisional interactions \cite{kaufman15} or using post-selected interference \cite{lester18}. Coupling qubits to Rydberg atoms overcomes this limitation by allowing strong, controllable long-range interactions between atoms \cite{jaksch00,lukin01,saffman10,saffman16}. For qubits separated below the blockade radius, typically 10~$\mu$m, the strong dipole-dipole interactions induce a conditional excitation process that allows a controlled phase-gate to be applied \cite{jaksch00}, as shown in Fig.~\ref{fig:entanglement}(b). This gate requires three laser pulses, first a $\pi$-pulse on the control qubit mapping $\vert 1 \rangle_c\rightarrow\vert r\rangle_c$ followed by a $2\pi$-pulse on the target qubit from $\vert 1 \rangle_t\rightarrow\vert r\rangle_t$. If the control qubit is in state $\vert r\rangle_c$ then blockade detunes this pulse from resonance on the target qubit and no phase is acquired, however if the control is in state $\vert 0\rangle_c$ then the target executes a complete rotation and acquires the phase shift of $e^{i\pi}$. A final pulse returns the control qubit to $\vert 1\rangle_c$ leading to the realisation of a controlled phase gate. Adding single qubit Hadamard gates to the control qubit before and after the gate allows conversion to a controlled-NOT gate, as first demonstrated by Isenhower \etal{}. \cite{isenhower10}.

Exploiting Rydberg states provides an extremely versatile interaction allowing controlled interactions between different isotopes \cite{zeng17} and atomic species \cite{beterov15}, suitable for embedding ancilla qubits that can be read out without perturbing the logical qubits for quantum error correction. Unlike other approaches, the long-range couplings are also ideal for performing multi-qubit gate operations including Toffoli \cite{isenhower11} and Deutsch gates \cite{shi17} allowing qubits to couple to all surrounding neighbours without the requirement of decomposition into sequential two-qubit gates. This approach enables high efficiency realisation of Grover's search algorithm \cite{molmer11,petrosyan16} and is capable of achieving fault-tolerant computing using surface codes \cite{auger17}.

The intrinsic limitations in Rydberg quantum gates arise from the finite Rydberg state lifetime and off-resonant excitation of non-blockaded Rydberg states \cite{saffman10,saffman16}. Operation at cryogenic temperatures allows suppression of black-body limited lifetimes \cite{beterov09}, whilst off-resonant excitation can be suppressed using shaped pulses based on the Derivative Removal by Adiabatic Gate (DRAG) technique \cite{motzoi09}. DRAG allows increased gate speed (minimising the error associated with finite Rydberg lifetime) while avoiding excitation of anti-blockaded states by tailoring the spectral properties of the light. A Rydberg-DRAG scheme using the single photon transition provides a 50~ns gate protocol with a theoretical fidelity of $F>0.9999$ at 300~K \cite{theis16}, competitive with speeds of solid-state qubits and well in excess of the threshold for fault-tolerant computing \cite{auger17}.

A major challenge in current experiments is to overcome the low-fidelity observed in previous demonstrations of entanglement generation using Rydberg atom interactions \cite{wilk10,isenhower10,zhang10,maller15,jau16,zeng17,picken18}. To date the highest published ground state entanglement fidelity is around 81\% using either Rydberg dressing \cite{jau16} or blockade \cite{picken18} after post-selection, well below the intrinsic gate error. Presently the dominant error is technical, arising from residual phase-noise of the laser system causing dephasing \cite{leseleuc18}. However, a route to circumvent this limitation has been demonstrated using a high-finesse cavity to filter unwanted frequency components from the light. This technique has enabled high-fidelity control of ground-Rydberg entanglement with a fidelity of $97\%$ \cite{levine18} comparable to solid-state approaches and paving the way to scalable universal quantum computation using Rydberg atom arrays.

As well as enabling digital computing, the Rydberg array platform also functions as a powerful analog quantum simulator for spin-models \cite{schauss18}. These provide a simplified many-body Hamiltonian describing highly-correlated condensed-matter systems enabling studies of out-of-equilibrium dynamics and quantum magnetism on system sizes greatly exceeding what can be simulated classically.

For atoms in the ground state $\vert g\rangle$ coupled to a Rydberg level $\vert r \rangle$ via a laser with detuning $\Delta$ and Rabi frequency $\Omega$, the system can be described by a transverse Ising Hamiltonian of the form \cite{labuhn16,bernien17}
\begin{equation}
\frac{\mathcal{H}}{\hbar} = \displaystyle\sum_i\frac{\Omega}{2}\sigma_x^i-\displaystyle\sum_i \Delta_in_i+\displaystyle\sum_{i\neq j}V_{ij}n_in_j,
\end{equation}
where $\sigma_x^i=\vert g_i \rangle\langle r_i \vert+\vert r_i\rangle\langle g_i\vert$ describes the coupling between ground and Rydberg levels, $n_i=\vert r_i \rangle\langle r_i\vert$ and $V_{ij}=C_6/R_{ij}^6$ is the van der Waals coupling between atoms in the array. Through coherent control of the laser coupling strength $\Omega_i$ and detuning $\Delta_i$ for each atom, atoms can be prepared into the many-body ground states of the strongly interacting systems allowing studies of quantum phase transitions and symmetry breaking, as well as studying out-of-equilibrium dynamics following quenches when rapidly changing system parameters. This has been implemented for a 1D chain of 51 qubits \cite{bernien17} and a variety of 2D geometries with up to 49 spins \cite{leseleuc18a}, allowing observation of transitions to ordered anti-ferromagnetic phases \cite{lienhard18} and experimental demonstration of the quantum Kibble-Zurek mechanism governing critical scalings associated with quantum phase transitions \cite{keesling19}.

Alternatively, by encoding qubits within the Rydberg manifold a microwave field can be used to drive resonant couplings between Rydberg states $\vert r\rangle$ and $\vert r'\rangle$, leading to interactions in the resonant dipole-dipole regime with highly anisotropic couplings of the form $C_3(\theta)/R^3_{ij}$ \cite{ravets15}. This mitigates technical limitations present in laser driving \cite{leseleuc19,levine18} and enables realisation of an XY or spin-exchange Hamiltonian of the form
\begin{equation}
\frac{\mathcal{H}}{\hbar} = \frac{1}{2}\displaystyle\sum_{i\neq j}V_{ij}(\theta)(\sigma_i^+\sigma_j^-+\sigma_i^-\sigma_j^+),
\end{equation}
where $\sigma^\pm_i = \sigma_i^x\pm i\sigma_i^y$ which has been used to study excitation transfer along a spin-chain \cite{barredo15} and excitation decay in an isolated atomic ensemble \cite{orioli18}. Exploiting the reconfigurable geometry and strongly anisotropic interactions to introduce asymmetric couplings, the XY Hamiltonian allows engineering of topologically-protected edge states that are robust to external perturbations as recently demonstrated for a 2D spin-chain \cite{leseleuc19a}. Extending this approach to more complex geometries, for example a honeycomb lattice, provides topological protection for robust quantum information storage \cite{weber18}.

Beyond studies of correlated many-body systems, this approach of programmable interactions is of direct relevance to quantum computation by exploiting the mapping of combinatorially hard optimisation problems onto the ground state of Ising-type spin Hamiltonians \cite{andrews14a}. This approach requires a system offering locally programmable spin-couplings as demonstrated using Rydberg arrays, and a robust algorithm for preparing atoms in the ground state through either quantum annealing \cite{farhi01,lechner15,glaetzle17} or variational quantum approximate optimisation algorithms (QAOA) \cite{farhi14} using a classical optimiser to drive the parameters controlling the quantum hardware.
Recent studies have shown the transverse Ising model above allows direct encoding of NP-complete problems using geometric arrangement of the atoms alone \cite{pichler18a}, enabling an efficient solution to the maximum independent set (MIS) problem even using current near-term hardware \cite{pichler18}.


\section{Electromagnetic Field Sensing with Rydberg States}
The use of atoms and atom-like systems for sensing is advantageous because many kinds of atoms are stable quantum objects. The properties of atoms can also be known very accurately by making precision measurements in specialised laboratories. In this sense, atoms can be used to make self-calibrated sensors because these properties are fixed for all atoms of the same isotopic species. Stability and self-calibration are advantageous because measurements can be traced back to precision measurements of the atomic characteristics and fundamental constants. Unlike man-made sensors, atoms are inherently stable and always the same so issues associated with manufacturing variations and aging can be eliminated, in principle. Rydberg atoms also have a large range of transition frequencies that can be exploited for conversion of hard to detect and generate frequencies to those which are more conventional, such as visible light.

Because the Coulomb potential supports an infinite number of electronic states, there are an infinite number of Rydberg transitions. In alkali atoms, these transitions span over 6 orders of magnitude in frequency, ranging from less than $1\,$MHz to approximately $1\,$THz. For many of these transitions, the dipole matrix elements are large, some of which are $100-1000$ times that of the D2 transition in alkali atoms. These properties make Rydberg atoms sensitive to electromagnetic radiation throughout the frequency range of their transitions, most notably from the microwave (MW) to terahertz regions of the electromagnetic spectrum. Although the sensitivity to electromagnetic fields has been understood for a long time \cite{gallagher05}, it was only recently appreciated that, coupled with atoms in small vapor cells \cite{kubler10} and sub-Doppler spectroscopy \cite{fleischhauer2005,mohapatra07}, Rydberg levels can be a useful atom-based electromagnetic field sensing technology \cite{sedlacek12,fan15}. The fact that the atoms are contained in a dielectric vapor cell is important because the electromagnetic field can be perturbed less than in the presence of a metal antenna \cite{fan15a}. Rydberg atom electromagnetic field sensing qualifies as a quantum technology in the broader sense of the concept. Many of the advantages of the method are rooted in the fact that the atom is a quantum object. As part of a broader theme, these types of sensors use weakly bound (trapped) electrons. Such systems possess exaggerated properties that can be easily manipulated and addressed. Other examples of related physics can be found in Refs. \cite{sedlacek16,quantumdots,Wrachtrup2006,tilman,Obrien2009}.

\begin{figure}[ht]
\begin{center}
\includegraphics[width=14cm]{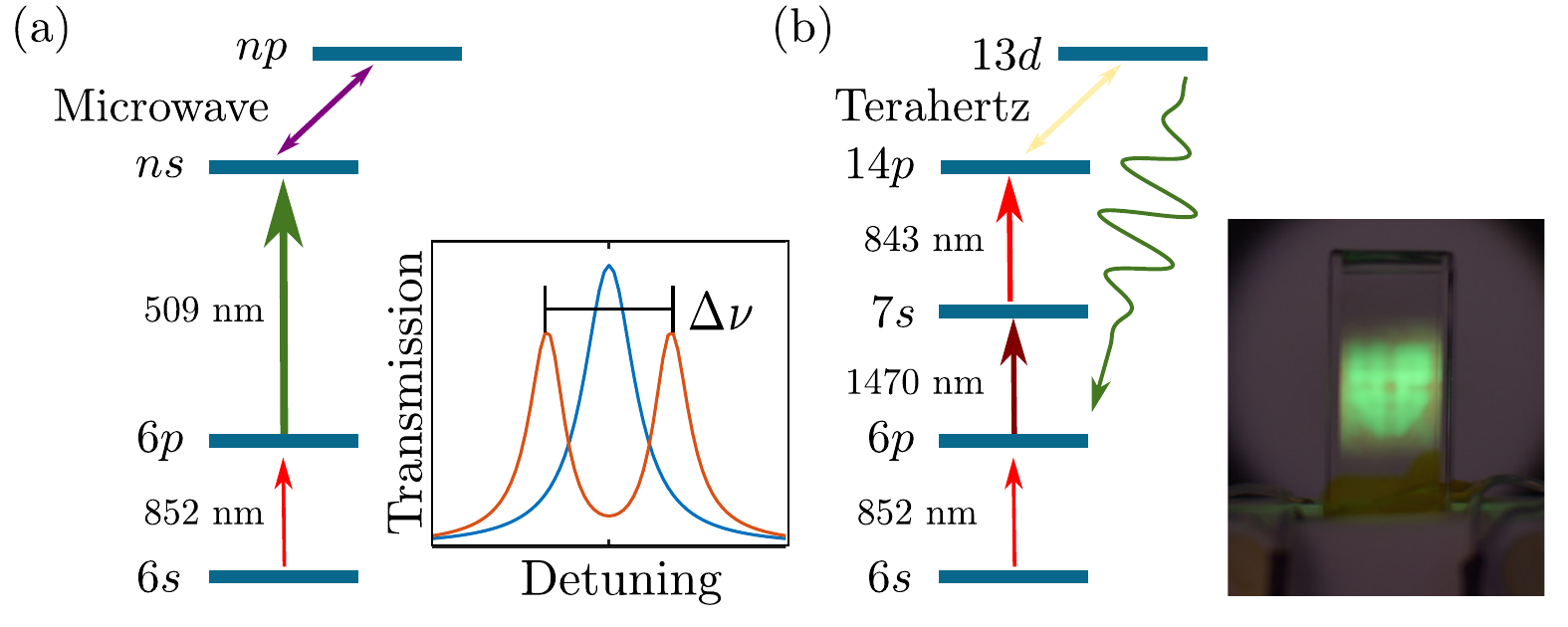}
\end{center}
\caption{Level schemes for microwave electric field sensing and terahertz imaging. (a) A typical laser configuration for microwave electric field sensing using cesium. A probe laser at $852\,$nm  and a coupling laser at $509\,$nm are used to setup electromagnetically induced transparency (EIT). The transmission of the probe laser is used to determine the effect that a resonant or quasi-resonant microwave field has on the upper state of the EIT system, which for strong fields appears as a splitting $\Delta\nu$ as shown in the inset. The concept can be generalised to use more laser fields as referenced in the text in order to overcome technical challenges such as the Doppler effect caused by the motion of the atoms in the vapor cell used as the sensor.  (b) For the terahertz imaging scheme described in the text, several laser transitions are used to excite a low lying Rydberg state. The terahertz radiation couples two Rydberg states which leads to fluorescence which is detected and used to image the terahertz radiation. Inset: Example image of a terahertz field transmitted through an aperture with the shape of a shield. Courtesy of Lucy Downes, Durham University.}
\label{fig:qtechlevels}
\end{figure}

Atom-based electromagnetic field sensors that utilise Rydberg states, as introduced here, operate at room temperature and use alkali atoms contained in a vapor cell to sense electromagnetic fields \cite{sedlacek12,fan15,Hollowayrev,gordon2014millimeter,wade16}. The two principal methods used for electromagnetic field sensing to date are shown in Fig.~\ref{fig:qtechlevels}. Light is used to create a system that has a Rydberg state component that is in resonance, or near-resonance, with the electromagnetic field that is targeted for detection. Either the transmission of a probe laser is monitored to determine the electromagnetic field, Fig~\ref{fig:qtechlevels}(a), or the fluorescence from a Rydberg state is collected to sense the presence of the electromagnetic field, Fig.~\ref{fig:qtechlevels}(b).

For the first of these methods, which is reviewed in Ref.~\cite{fan15}, Fig~\ref{fig:qtechlevels}(a), the light creates a state of the atom that is a quantum interference via electromagnetically induced transparency (EIT) that is sensitive to an incident high frequency (MHz-THz) electric field via the dipole coupling of the electromagnetic field and the atomic state. The response induced in the atom is read-out by observing the absorption of a probe laser beam passing through a vapor cell. The atoms in the vapor cell are put into a state that acts as a coherent quantum interferometer with laser fields. The light from a resonant probe laser beam passes through a spectrally narrow, transparency window in a normally absorbing material, i.e. cesium or rubidium vapor, due to the presence of the strong coupling laser beam. The probe laser energies where the light transmission takes place depend on the Rydberg state energy level positions.
If the field is strong enough to split the Rydberg state, two probe transmission peaks are observed with a frequency splitting corresponding to the Rabi frequency of the target electromagnetic fields interaction with the Rydberg transition,
\begin{equation}
\Delta \nu = \frac{\vec{E}_0 \cdot \vec{\mu}}{h}.\label{splitting}
\end{equation}
$\Delta \nu$ is the frequency interval between the transmission peaks caused by the MW electric field. $\vec{E}_0$ is the electric field vector and $\vec{\mu}$ is the corresponding transmission dipole moment for the Rydberg transition. $h$ is Planck's constant. One only needs to know the transition dipole moment, a property of the atom that can be measured using precision spectroscopy, to determine the electromagnetic field amplitude. The range of MW electric fields that can be measured can be extended beyond the Autler-Townes regime by measuring the MW electric field induced change in transmission when the coupling and probe lasers are tuned to resonance \cite{sedlacek12}.  The sensitivity of the method as used to date is determined by shot noise in the probe laser; dephasing mechanisms, such as collisions, transit time broadening and blackbody decay; and the residual Doppler shifts of the EIT scheme. Photon shot-noise limited performance has been achieved \cite{kumar17,kumar17a}. The residual Doppler shifts can be overcome using a 3-photon scheme in cesium \cite{kubler18}. The limitations of photon shot noise have prevented the projection noise limit, which can reach better than pV$\,$cm$^{-1}$Hz$^{-1/2}$ depending on geometry, from being realised. The accuracy is limited by several factors including the vapor cell material and geometry \cite{fan15a} and how well the transition dipole moments are known, i.e. the atomic spectrum. For small fields, where changes in amplitude of the probe transmission signal are used, the Rabi frequency of the coupling laser must be known in order for the absolute field to be determined. To date, amplitude modulation \cite{sedlacek12}, frequency modulation \cite{kumar17}, homodyne \cite{kumar17a},  dispersive detection \cite{fan16} and heterodyne \cite{shaffer18,jing19,holloway19b} schemes for detecting the light signal have been demonstrated. Frequency modulation, homodyne and heterodyne detection all achieved photon shot-noise, or near photon shot noise, limited performance for realistic measurement parameters. These works suggested that the sensor is photon shot noise limited.

The scheme for terahertz imaging using atomic fluorescence is shown in Fig~\ref{fig:qtechlevels}(b). The approach described in Refs.~\cite{wade16,wade19} uses Rydberg transitions to convert a terahertz electromagnetic field into light, which can be imaged by more conventional detectors. After an initial Rydberg state has been prepared, a resonant terahertz electromagnetic field excites the system to another Rydberg state. The upper Rydberg state can decay by spontaneous emission, which creates light that can be imaged with optical resolution and detected with straightforward silicon based photodetectors. The lower Rydberg state can also fluoresce, but it primarily creates photons that decay to the ground state, since it is a p-state. Choosing the upper state as a d-state leads predominantly to decay into the lowest p-state. The difference between the color of light emitted by the 2 transitions enables the upper state fluorescence to be distinguished so that it can be used to detect the terahertz field. Although this method is not self-calibrated, it can be a useful method for detecting terahertz radiation which is difficult to do. A number of other terahertz detection technologies exist, but it has been shown that this one can reach image acquisition rates of $\sim 3\,$kHz \cite{wade19}. The signal strength is determined by the Rabi frequency of the terahertz electromagnetic field interacting with the atom; the decay rate into the detection channel compared to competing processes, like blackbody decay; and the fluorescence collection efficiency.

Minimum detectable electric field amplitudes of $< 1\,\mu$V$\,$cm$^{-1}$ with a sensitivity of $\sim\ 1\,\mu$V$\,$cm$^{-1}\,$Hz$^{-1/2}$ and accuracy of $< 1\%$ have been demonstrated for Rydberg atom-based electric field sensing. The exact numbers depend on the geometry of the vapor cell and the Rabi frequencies of the laser fields. Microwave electric fields in the mm-wave regime have been measured \cite{gordon2014millimeter}. The method can be sensitive to the vector electric field \cite{sedlacek13}. Phase retrieval using Rydberg atoms has been investigated in both imaging \cite{shaffer18} and single microwave electromagnetic field point detection \cite{jing19,holloway19b}. Imaging of electric fields near MW devices was first demonstrated in Ref.~\cite{fan14}. These results produced sub-wavelength imaging of the target fields determined by the optical imaging resolution. Similarly, imaging inside a vapor cell was demonstrated in Ref.~\cite{holloway14}. Strong field detection has been investigated in Ref.~\cite{anderson16}. Rydberg atoms have also been studied as a receiver, first in Ref.~\cite{fatemi18a}, including an interesting follow-up study on the 'smallness' of the atom as an antenna \cite{fatemi18b}. Several recent works on receivers have appeared recently \cite{otago18,song19,anderson18b}. A guitar was even recorded using the atoms in the vapor cell \cite{holloway19g}. For terahertz imaging, the sensitivity that has been demonstrated is $\sim 200\,$fW$\,$Hz$^{-1/2}$ \cite{wade19}. There has been less work done on the terahertz imaging scheme described here.

\section{Rydberg Quantum Optics}

In Rydberg quantum optics \cite{firstenberg16}, the goal is to exploit the strongly-interacting properties of Rydberg atoms to control light at the level of single photons, examples include single photon sources, single photon transistors, and photonic phase gates. In this section we shall discuss some of these applications, but first we introduce some physics concepts that allow us to understand how Rydberg interactions affect light propagation inside a medium. We are interested in how light, at the few quanta level, propagates through a medium. We can assume that the medium is an ensemble of two-level atoms with ground and excited states labelled $\vert 0 \rangle$ and $\vert 1\rangle$, respectively. A single photon localised inside the medium at time $t=0$ can be described by a polaritonic wave function of the form
\begin{eqnarray}
\vert \psi\rangle =\frac{1}{\sqrt{{\cal N}}}
\sum_j c_j {\rm e}^{{\rm i}\boldsymbol{k}\cdot\boldsymbol{r}_j}\vert 0\ldots 0 1_j 0\ldots 0\rangle~,\label{eq:polariton}
\end{eqnarray}
where ${\cal N}$ is the number of atoms, $\boldsymbol{r}_j$ is the position of atom $j$, and $c_j$ is an amplitude coefficient which will also depend on time and the position of the atom. The polariton wave function states that one atom is excited, because there is one photon inside the medium. However, we do not know and we cannot know which atom is excited. If we did know, then the state would be different from the one written in eqn.~\ref{eq:polariton}. The important parameters of the polariton are the amplitudes, $c_j$, and the phase terms ${\rm e}^{{\rm i}\boldsymbol{k}\cdot\boldsymbol{r}_j}$ as these determine the spatial mode and ensure that the photon exiting the medium is the same as the one that entered. If the phases evolve,  ${\rm e}^{{\rm i}\boldsymbol{k}\cdot\boldsymbol{r}_j}\rightarrow {\rm e}^{{\rm i}\boldsymbol{k}\cdot\boldsymbol{r}_j+\phi_j}$ in a way that is different for different atoms then the photon leaving the medium is in a different mode, i.e., scattered. For example, if the atoms move due to finite temperature this causes a motional-induced dephasing of the polariton, and the photon is scattered. For room temperature atoms this dephasing occurs on a time scale of order nanoseconds \cite{whiting17}.

In examples of multi-level light-atom interactions like EIT, two-photon excitation or four-wave mixing, the excited state $\vert 1\rangle$ is typically coupled to another state $\vert 2\rangle$ using a classical control field with Rabi frequency $\Omega_{\rm c}$.  In the case of EIT, the polariton becomes a dark-state polariton where the $\vert 1\rangle$ in eqn~(\ref{eq:polariton}) can be replaced by $\vert 2\rangle$ \cite{fleischhauer00}.
For a coupling Rabi frequency greater than the excited state decay rate, $\Omega_{\rm c}\gg \Gamma$, the medium is transparent to light resonant with the $\vert 0\rangle$ to $\vert 1\rangle$ transition and the dark-state polariton is equivalent to a photon that propagates at the speed of light.  For  $\Omega_{\rm c}\sim \Gamma$, the medium is still transparent but now the dark-state polariton is a mixture of photon and atomic excitation and propagates at a speed equal to $c/n_{\rm g}$, where the group index depends on the slope of the dispersion and can be very large $>10^6$ giving rise to slow light \cite{fleischhauer00}. Reducing $\Omega_{\rm c}$ narrows the width of the transparency window and increases the group index. In the limit $\Omega_{\rm c}\rightarrow0$  the dark-state polariton becomes a purely atomic excitation and is stored inside the medium. If the control field  is turned back on then the stored polariton is converted back into a photon and exits the medium. Using the control field to write and read an excitation is the basis of some quantum memory schemes.

In  Rydberg quantum optics, a ladder-type excitation scheme is used \cite{mohapatra07,firstenberg16}, similar to Fig.~\ref{fig:qtechlevels}. In this case the excited state $\vert 1\rangle$ is coupled to a highly-excited Rydberg state $\vert r\rangle$, and we can replace the $\vert 1\rangle$ in eqn~(\ref{eq:polariton}) by $\vert r\rangle$ and refer to this state as a Rydberg polariton. As the polariton contains Rydberg character it has similar strongly-interacting properties to individual Rydberg atoms.
For example, there is a blockade effect preventing the excitation of two Rydberg polaritons within a blockade radius. Consequently if the ensemble is smaller than the blockade volume then only one excitation is possible. By reading out this excitation, by turning on the control field, it is possible to generate a single photon. Below we discuss two experiments that exploit this idea.

\begin{figure}[tb]
\begin{center}
\includegraphics[width=7.5 cm]{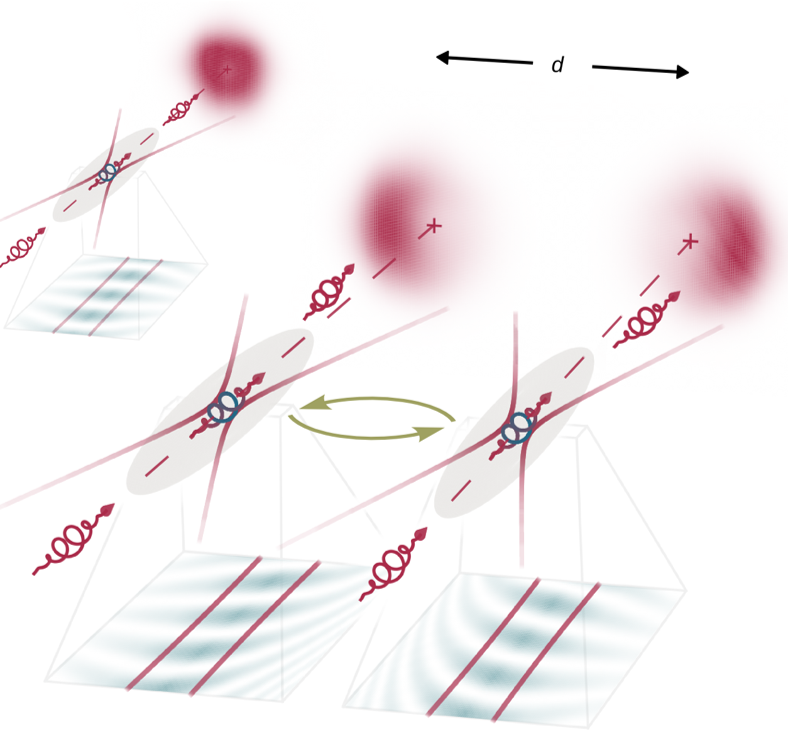}
\end{center}
\caption{The effect on an interaction-induced phase shift on the propagation of photons. The inset shows a single polariton propagation through a medium. The phase is preserved and the output mode matches the input mode. The main figure shows two neighbouring polaritons which interact via a repulsive van der Waals potential. In this case the interaction-induced phase change leads to a deflection of the outgoing photons. Figure courtesy of Paul Huillery. }
\label{fig:deflect}
\end{figure}

 In applications such as a photonic phase gate we would like a controlled phase shift, e.g. $\phi_j=\pi$, that is the same for all atoms. This can be achieved experimentally using blockade \cite{tiarks19}. However, in addition to blockade, there is also an interaction-induced phase shift \cite{bariani12} which can give rise to non-uniform phase shifts. Consider two ensembles side by side as shown in Fig.~\ref{fig:deflect}. Two photons entering the two ensembles create two Rydberg polaritons, initially with a uniform phase gradient, like the single channel case shown in the inset. The combined state is given by the product of two individual polaritons. The Rydberg components interact via a van der Waals interaction, $V$, such that terms involving atom $i$ in ensemble 1 and atom $j$ in ensemble 2 pick up a phase $V_{ij}t$ where $t$ is the interaction time. As the interaction potential is proportional to $1/r_{ij}^6$, where $r_{ij}$ is the distance between atoms $i$ and $j$, the phase shift is larger for closer parts of the two ensembles. The interaction-induced phase shift is shown in the projections below the ensembles  in  Fig.~\ref{fig:deflect}. In this example, the non-uniform phase shift leads to a deflection of the outgoing photons---as if the photon were a particle moving in the potential created by the van der Waal interaction of another photon. The effective force between the photons may be either repulsive or attractive depending on the sign of the van der Waals shift.

If the two channels in Fig.~\ref{fig:deflect} are moved closer then the phase shifts increase. If they are overlapped then the phase shift of the doubly-excited polariton mode becomes so distorted that there is very little overlap between the outgoing two-photon mode and the unperturbed mode. In this case, any two, or more, photon components are completely dephased and scattered out of the forward propagation mode. For a coherent state input, only the zero and one photon components survive leading to strong anti-bunching of the forward transmitted light. This anti-bunching effect was first demonstrated in a series of experiments in 2012 \cite{dudin12,peyronel12,maxwell13}. The two-channel photon deflection effect illustrated in Fig.~\ref{fig:deflect} was demonstrated by Busche {\it et al}. in 2017 \cite{busche17}. This interaction-induced dephasing effect can be used to realise a single-photon source.

\label{sec:single_photon_source}

\begin{figure}[tb]
\begin{center}
\includegraphics[width=8.6 cm]{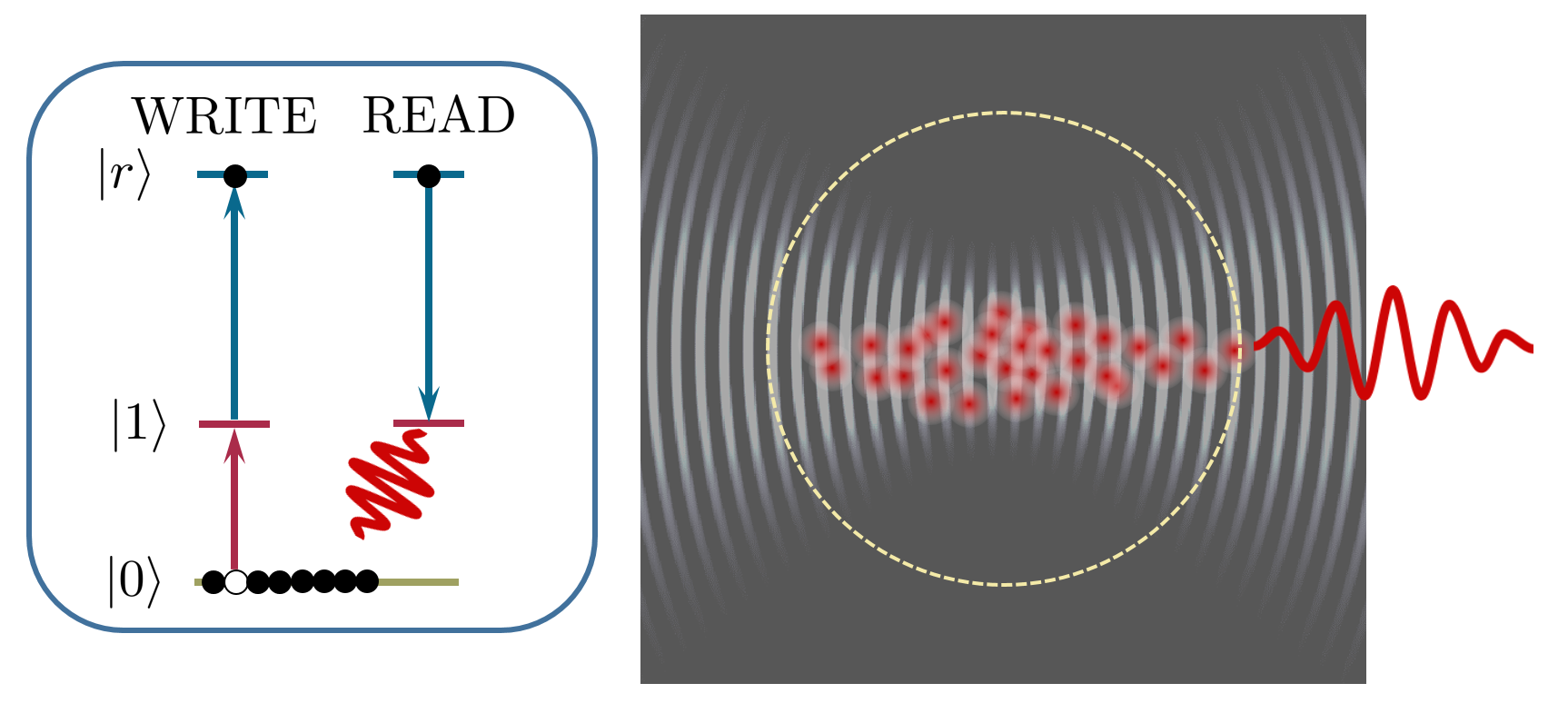}
\end{center}
\caption{Principle of a Rydberg single-photon source: A single excitation is excited to a highly-excited Rydberg state $\vert r\rangle$ using a two-photon transition as shown on the left. This writes a Rydberg polariton into the medium with a phase pattern determined by the wave vector of the excitation light, shown schematically on the right. A read pulse maps the polariton onto the intermediate state $\vert 1\rangle$ which emits a photon in a direction determined by the polariton phase pattern, see also Fig.~\ref{fig:deflect}.  }
\label{fig:single_photon_source}
\end{figure}

A wide variety of single-photon sources has been demonstrated ranging from heralded photons produced in parametric down-conversion to four-wave mixing to near deterministic sources based on excitation of single emitters such as single atoms and quantum dots \cite{darquie05,senellart17}. The principle of a Rydberg atom-based single-photon source is relatively straightforward \cite{saffman02}, however, a full theoretical treatment requires solving a complex quantum many-body problem. Consider a two-photon excitation to a Rydberg state in an atomic ensemble with size less than the blockade volume such that only one polariton is excited, as illustrated in Fig.~\ref{fig:single_photon_source}. Subsequently the polariton is read-out using the coupling laser to transfer the excitation into the excited state $\vert 1\rangle$ which emits a photon in a direction determined by the phase pattern. This type of single-photon source was demonstrated using laser-cooled atoms in an optical dipole trap by Dudin an Kuzmich in 2012  \cite{dudin12}. In practice, the interaction-induced dephasing effect discussed above can be as important as blockade in determining the performance of the source. In principle, it is possible to implement the above scheme using thermal atoms in a room temperature vapour, however, in this case the motional dephasing is much faster and the coherence time is on the order of nanoseconds, consequently Rabi frequencies in the gigahertz range are needed \cite{Ripka2018}.

\label{sec:photonic_phase_gate}

In addition to single-photon sources and detectors, the third basic building block of an all-optical quantum network is a photon switch or photon-photon gate, where a `control` photon either redirects or shifts the phase of a `target` photons. A variant of a photon switch is a photon transistor where the control photon modifies the 'flow' of a target photons similar to how the gate in an electronic transition controls the flow of current between the source and drain. All-optical photonic transistor using Rydberg blockade where first demonstrated in 2014 \cite{tiarks16,gorniaczyk16}.

\begin{figure}[tb]
\begin{center}
\includegraphics[width=8.6 cm]{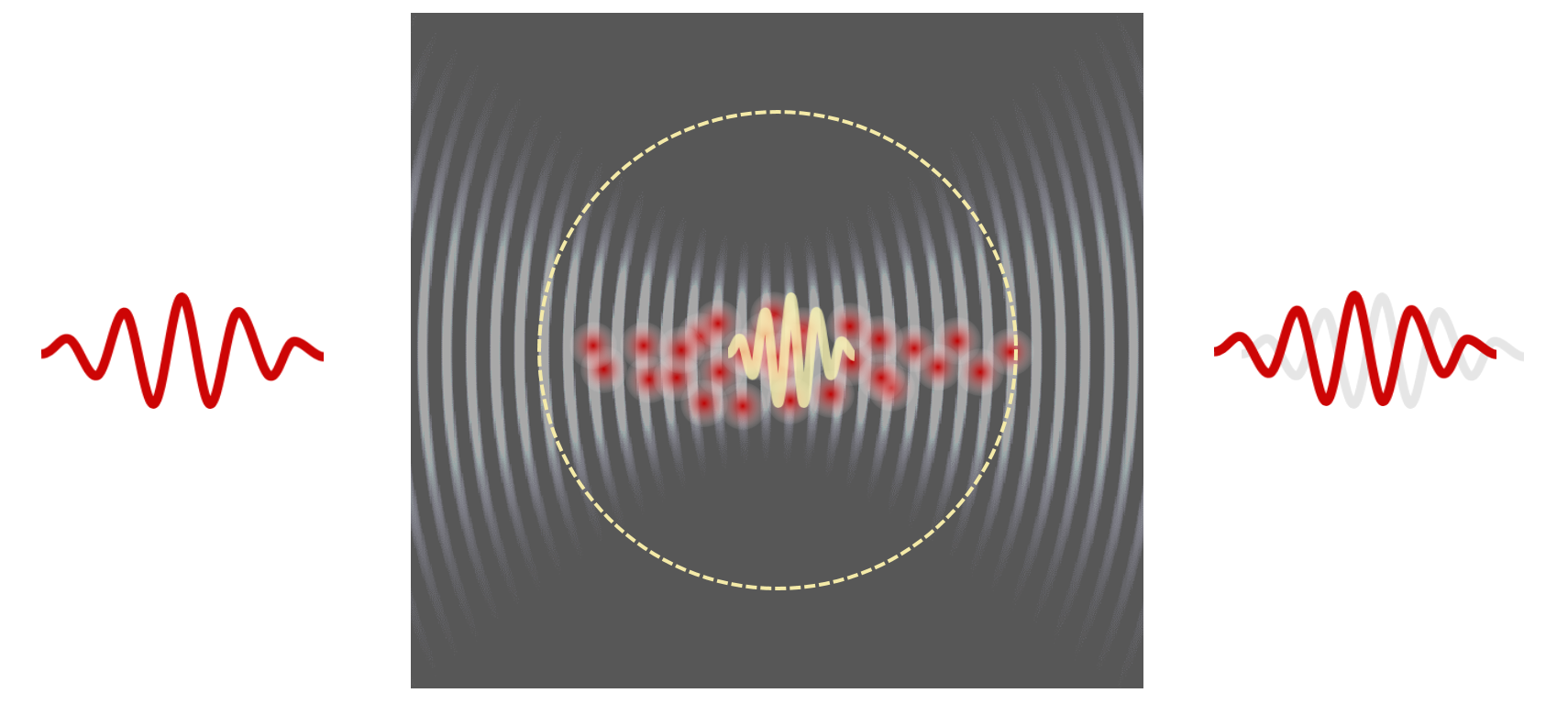}
\end{center}
\caption{Principle of the photon-photon gate reported in \cite{tiarks19}: First a control photon is localised inside the medium as a Rydberg polariton (indicated by the yellow line). Next, a target photon (red line) is sent through the same medium with a coupling to another Rydberg state such that the control and target interact via a dipole-dipole interaction leading to a $\pi$ phase shift. The phase of the outgoing target photon when the control photon is not present is shown in grey.  }\label{fig:photon_phase_gate}
\end{figure}

In linear optics, it is possible to implement photon-photon gates probabilistically using the Hong-Ou-Mandel effect \cite{knill01}. Although deterministic photonic phase gates based on Rydberg EIT were proposed theoretically in 2005 \cite{friedler05}, it was thought that there ``no known or foreseen material has an optical non-linearity strong enough to implement this [i.e. $\pi$] conditional phase shift'' \cite{obrien07}. Experimentally, the prospects for deterministic photonic phase gates were significantly advanced, first by the demonstration of giant optical non-linearities in 2010 \cite{pritchard10}, and second with the demonstration of anti-bunching using Rydberg EIT in 2012 \cite{peyronel12}. In principle, there are many ways that photon-photon gates can be realised, see e.g. \cite{friedler05,gorshkov11,Paredes14,Khazali15}. The general principle is convert the incoming photons into polaritons using EIT and subsequently use blockade to implement a condition phase shift similar to the Rydberg atom gate \cite{jaksch00}. The first experimental demonstration of a photonic phase gate with a $\pi$ phase shift based on Rydberg EIT was reported in 2019 \cite{tiarks19}. The scheme is illustrated schematic in Fig.~\ref{fig:photon_phase_gate}. A control photon is localised inside the medium as a Rydberg polariton (indicated by the yellow line). Next, a target photon (red line) is sent through the same medium with a coupling to another Rydberg state such that the control and target interact via a dipole-dipole interaction leading to a $\pi$ phase shift.

In summary, Rydberg quantum optics has made tremendous advances in only ten years, and Rydberg ensembles have become established as the only known medium with a non-linearity sufficiently large to not require a cavity in order to implement an all-optical quantum gate. In future, research is likely to focus on interfacing the deterministic single-photon sources discussed above with quantum gates and repeater protocols to realise an all-optical quantum network, and provide quantum communication channels between Rydberg quantum computers.

\section{Outlook}

Rydberg series are observed in any system with bound electrons, including atoms, ions, molecules or excitons in semiconductors, but in this review, we have mainly focused on single-electron Rydberg atoms. As the properties of Rydberg atoms scale with the principle quantum number, it is possible to engineer quantum systems with controllable atomic interactions that couple strongly to either optical, terahertz or microwave fields or combinations thereof. Consequently, Rydberg atoms offer a highly-versatile, reproducible, and tunable component that can be used as the basis for a variety of quantum technologies. Particularly important currently are the applications of Rydberg atoms in quantum computing, sensing and imaging, and quantum optics. The use of Rydberg atoms for quantum technology is currently evolving at a faster pace than ever, and it may be dangerous to speculate on what might happen next.  However we can be sure that as the field matures, more and more of the concepts outlined above will be translated from laboratory demonstrators into commercial systems that will increasingly help to solve a wide range of societal problems.

\section*{Acknowledgments}
CSA acknowledges financial support from EPSRC Grant Ref. Nos. EP/M014398/1, EP/P012000/1, EP/R002061/1, EP/R035482/1, and EP/S015973/1. JDP acknowledges financial support from EPSRC Grant Ref. Nos. EP/N003527/1 and EP/S015973/1. JPS acknowledges financial support from the Air Force Office of Scientific Research grant FA9550-12-1-0282 and the National Science Foundation research grant PHY-1607296.
\section*{References}

\providecommand{\newblock}{}

\end{document}